\def\ra{\rangle}
\def\la{\langle}

\def\be{\begin{equation}}
\def\ee{\end{equation}}
\def\ba{\begin{array}}
\def\ea{\end{array}}

\def\ra{\rangle}
\def\la{\langle}

\documentstyle[11pt]{article}
\input amssym.def
\begin{document}
\baselineskip=18pt \setcounter{page}{1}
\begin{center}
{\LARGE \bf Canonical Form and  Separability of PPT States in
${\cal C}^2 \otimes {\cal C}^M \otimes {\cal C}^N$ Composite
Quantum Systems}
\end{center}
\vskip 2mm

\begin{center}
{\normalsize Xiao-Hong Wang$^1$, Shao-Ming Fei$^{1,\  2}$,
  Zhi-Xi Wang$^1$, and Ke Wu$^1$  }
\medskip

\begin{minipage}{4.8in}
{\small \sl $ ^1$
Department of Mathematics, Capital  Normal University, Beijing,
China.}\\
{\small \sl $ ^2$ Institute of Applied Mathematics, University of
Bonn,  53115 Bonn, Germany}
\end{minipage}
\end{center}

\vskip 2mm
\begin{center}

\begin{minipage}{4.8in}

\centerline{\bf Abstract}
\bigskip

We investigate the canonical forms of positive partial
transposition (PPT) density matrices in ${\cal C}^2 \otimes {\cal
C}^M \otimes {\cal C}^N$ composite quantum systems with rank $N$.
A general expression for these PPT states are explicitly obtained.
From this canonical form a sufficient separability condition is
presented.

\vskip 1mm
PACS number(s): 03.67.Hk, 03.65.Ta, 89.70.+c
\end{minipage}
\end{center}

\vskip0.4cm

Quantum entangled states have become one of the key resources in
the rapidly expanding field of quantum information processing and
computation \cite{DiVincenzo,teleport,dense,crypto}. Nevertheless,
the study of physical character and mathematical structure of the quantum
entanglement is far from being satisfied. One even does not
have a general criterion to judge if a quantum (mixed) state is
entangled or not. For bipartite states a number of entanglement measures such
as entanglement of formation and distillation \cite{568,7}, negativity
\cite{Peres96a}, von Neumann entropy and relative
entropy \cite{568,sw} have been proposed.
However most proposed measures of entanglement involve
extremizations which are difficult to handle analytically. For
instance, explicit analytic formulae for
entanglement of formation \cite{17} have been found only for
a pair of qubits system \cite{HillWootters}, and for some symmetric
states \cite{th} and a class of special states \cite{ent3}.
For multipartite systems there is no well
defined measure of entanglement yet.

The separability problem for pure states is quite well
understood \cite{peresbook}.
Nevertheless, in real conditions, due to the interactions with
environment, one encounters mixed states rather than pure ones.
The manifestations of mixed-state entanglement can be very subtle \cite{10}.
The Bell inequalities satisfied by a
separable system give the first necessary condition for
separability \cite{Bell64}. Afterwards the Peres criterion
\cite{Peres96a} says that partial transpositions with respect to
one or more subsystems of a separable state $\rho$ are positive.
This criterion
was further shown to be also sufficient for bipartite systems in
${\cal C}^2\otimes {\cal C}^2$ and ${\cal C}^2\otimes {\cal C}^3$
\cite{3hPLA223}. The reduction criterion proposed independently in
\cite{2hPRA99} and \cite{cag99} gives another necessary criterion
which is equivalent to the Peres criterion for ${\cal C}^2\otimes
{\cal C}^N$ composite systems but is generally weaker. There are
many other necessary criteria such as majorization
\cite{nielson01}, entanglement witnesses
\cite{3hPLA223,ter00}, extension of Peres criterion \cite{dps}, matrix
realignment \cite{ru02}, generalized partial transposition
criterion (GPT) \cite{chenPLA02}, generalized reduced criterion
\cite{grc}. For low rank density matrices there are also some
necessary and sufficient criteria of separability
\cite{hlpre}.

The separability and entanglement in ${\cal C}^2\otimes {\cal
C}^2\otimes {\cal C}^N$ and ${\cal C}^2\otimes {\cal C}^3\otimes
{\cal C}^N$ composite quantum systems have been studied in terms
of matrix analysis on tensor spaces \cite{22n}. It is shown that
all such quantum states $\rho$ with positive partial
transpositions and rank $r(\rho)\leq N$ are separable. The
canonical form and a sufficient separable condition PPT states in
${\cal C}^2 \otimes {\cal C}^2\otimes {\cal C}^2 \otimes {\cal
C}^N$ with rank $N$ are given in \cite{222n}. In this article we
extend the results in \cite{22n} to the case of composite quantum
systems in ${\cal C}^2 \otimes {\cal C}^M\otimes {\cal C}^N$. We
give a canonical form of positive partial transposition (PPT)
states in ${\cal C}^2 \otimes {\cal C}^M\otimes {\cal C}^N$ with
rank $N$ and present a sufficient separability criterion.

A separable state in ${\cal C}^2_{\sc a}  \otimes {\cal C}^M_{\sc
b} \otimes {\cal C}^N_{\sc c}$ is of the form:
\begin{equation}
\rho_{\sc abc}=\sum_{i}p_{i}\rho^{i}_{\sc a}\otimes \rho^{i}_{\sc
b} \otimes \rho^{i}_{\sc c}, \label{sep}
\end{equation}
where $\sum_{i}p_{i}=1$, $0< p_i\leq 1$, $\rho^{i}_{\alpha}$ are
density matrices associated with the subsystems $\alpha$,
$\alpha={\sc a,b,c}$. In the following we denote by $R(\rho)$,
$K(\rho)$, $r(\rho)$ and $k(\rho)$ the range, kernel, rank,
dimension of the kernel of $\rho$, respectively.

We first derive a canonical form of PPT states in ${\cal C}^2_{\sc
a} \otimes {\cal C}^4_{\sc b} \otimes {\cal C}^N_{\sc c}$ with
rank $N$, which allows for an explicit decomposition of a given
state in terms of convex sum of projectors on product vectors. Let
$|0_A\ra$, $|1_A\ra$; $|0_B\ra$, $|1_B\ra$, $|2_B\ra$, $|3_B\ra$
and $|0_C\ra\,, \cdots \,,|N-1_C\ra$ be some local bases of the
sub-systems ${\sc a}$, ${\sc b}$ and ${\sc c}$ respectively.

{\bf Lemma 1. }\ \  Every PPT state $\rho$ in ${\cal C}^2_{\sc a}
\otimes {\cal C}^4_{\sc b}\otimes {\cal C}^N_{\sc c}$ such that
$r(\la 1_A, 3_B|\rho |1_A, 3_B\ra)=r(\rho)=N$, can be transformed
into the following canonical form by using a reversible local
operation: \be \rho=\sqrt{F}[DC\ \ DB\ \ DA\ \ D \ \ C\ \ B\ \ A\
\ I]^{\dag} [DC\ \ DB\ \ DA\ \ D \ \ C\ \ B\ \ A\ \ I]\sqrt{F},
\ee where $A$, $B$, $C$, $D$, $F$ and the identity $I$ are
$N\times N$ matrices acting on ${\cal C}_{\sc c}^N$ and satisfy
the following relations: $[A,\ A^{\dag}]=[B,\ B^{\dag}]=[C,\
C^{\dag}]=[D,\ D^{\dag}]=[B,\ A]=[B,\ A^{\dag}] =[C,\ A]=[C,\
A^{\dag}]=[D,\ A]=[D,\ A^{\dag}]=[C,\ B]=[C,\ B^{\dag}]=[D,\
B]=[D,\ B^{\dag}]= [D,\ C]=[D,\ C^{\dag}]=0$ and $F=F^{\dag}$
($\dag$ stands for the transposition and conjugate).

{\bf Proof.}\ In the considered basis a density matrix $\rho$ can
be always written as: \be\label{r} \rho=\left(
\begin{array}{cccccccc}
E_1&E_{12}&E_{13}&E_{14}&E_{15}&E_{16}&E_{17}&E_{18}\\
E_{12}^{\dag}&E_2&E_{23}&E_{24}&E_{25}&E_{26}&E_{27}&E_{28}\\
E_{13}^{\dag}&E_{23}^{\dag}&E_{3}&E_{34}&E_{35}&E_{36}&E_{37}&E_{38}\\
E_{14}^{\dag}&E_{24}^{\dag}&E_{34}^{\dag}&E_4&E_{45}&E_{46}&E_{47}&E_{48}\\
E_{15}^{\dag}&E_{25}^{\dag}&E_{35}^{\dag}&E_{45}^{\dag}&E_5&E_{56}&E_{57}&E_{58}\\
E_{16}^{\dag}&E_{26}^{\dag}&E_{36}^{\dag}&E_{46}^{\dag}&E_{56}^{\dag}&E_6&E_{67}&E_{68}\\
E_{17}^{\dag}&E_{27}^{\dag}&E_{37}^{\dag}&E_{47}^{\dag}&E_{57}^{\dag}&E_{67}^{\dag}&E_7&E_{78}\\
E_{18}^{\dag}&E_{28}^{\dag}&E_{38}^{\dag}&E_{48}^{\dag}&E_{58}^{\dag}&E_{68}^{\dag}&E_{78}^{\dag}&E_8\\
\end{array}
\right),
\ee
where $E's$ are $N \times N$ matrices, $r(E_8)=N$.
After the projection $\tilde{\rho}=\la 1_A|\rho|1_A\ra$, we obtain
\be
\tilde{\rho}=\la1_A|\rho|1_A\ra
=\left(
\begin{array}{cccc}
E_5&E_{56}&E_{57}&E_{58}\\
E_{56}^{\dag}&E_6&E_{67}&E_{68}\\
E_{57}^{\dag}&E_{67}^{\dag}&E_7&E_{78}\\
E_{58}^{\dag}&E_{68}^{\dag}&E_{78}^{\dag}&E_8
\end{array}
\right).
\ee

$\tilde{\rho}$ is now a state in ${\cal C}^4_{\sc b} \otimes {\cal
C}^N_{\sc c}$ with $r(\tilde{\rho})=r(\rho)=N$. As every principal
minor determinant of $\tilde \rho^{t_B}$ ($\tilde \rho^{t_C}$) is
some principal minor determinant of $\rho$, the fact that $\rho$
is PPT implies that $\tilde \rho$ is also PPT, $\tilde{\rho}\ge
0$. After performing a reversible local non-unitary ``filtering"
$\frac{1}{\sqrt{E_8}}$ on ${\cal C}^N_{\sc c}$ and using the Lemma
5 in \cite{hlpre} we can express the matrix $\tilde{\rho}$ as
\be
\tilde{\rho}=\left(
\begin{array}{cccc}
C^{\dag}C&C^{\dag}B&C^{\dag}A&C^{\dag}\\
B^{\dag}C&B^{\dag}B&B^{\dag}A&B^{\dag}\\
A^{\dag}C&A^{\dag}B&A^{\dag}A&A^{\dag}\\ C&B&A&I
\end{array}
\right),
\ee
where
$[A,A^{\dag}]=[B,B^{\dag}]=[C,C^{\dag}]=[B,A]=[C,A]=[C,B]=[B,A^{\dag}]=[C,A^{\dag}]=[C,B^{\dag}]=0$.

Similarly, if we consider the projection $\la 3_B|\rho|3_B\ra$ ,
for the same reasons as above we conclude that the resulting
matrix
$$ \ba{rcl}
\bar{\rho}&=&\la 3_B|\rho| 3_B\ra\\[5mm]
&=&\left(
\begin{array}{cc}
E_4&E_{48}\\
E_{48}^{\dag}&E_8
\end{array}
\right)
=\left(
\begin{array}{cc}
D^{\dag}D&D^{\dag}\\D&I
\end{array}
\right)
\ea
$$
where $[D,D^{\dag}]=0$.

Hence, after performing the local filtering operation
$\frac{1}{\sqrt{E_8}}$, the matrix $\rho$ now has the form:
\be\label{rm}
\rho=\left(
\begin{array}{cccccccc}
E_1&E_{12}&E_{13}&E_{14}&E_{15}&E_{16}&E_{17}&E_{18}\\
E_{12}^{\dag}&E_2&E_{23}&E_{24}
&E_{25}&E_{26}&E_{27}&E_{28}\\
E_{13}^{\dag}&E_{23}^{\dag}&E_3&E_{34}
&E_{35}&E_{36}&E_{37}&E_{38}\\
E_{14}^{\dag}&E_{24}^{\dag}&E_{34}^{\dag}&D^{\dag}D&E_{45}
&E_{46}&E_{47}&D^{\dag}\\
E_{15}^{\dag}&E_{25}^{\dag}&E_{35}^{\dag}&E_{45}^{\dag}&C^{\dag}C
&C^{\dag}B&C^{\dag}A&C^{\dag}\\
E_{16}^{\dag}&E_{26}^{\dag}&E_{36}^{\dag}&E_{46}^{\dag}&
B^{\dag}C&B^{\dag}B&B^{\dag}A&B^{\dag}\\
E_{17}^{\dag}&E_{27}^{\dag}&E_{37}^{\dag}&E_{47}^{\dag}&
A^{\dag}C&A^{\dag}B&A^{\dag}A&A^{\dag}\\
E_{18}^{\dag}&E_{28}^{\dag}&E_{38}^{\dag}&D&C&B&A&I\\
\end{array}
\right).
\ee
Owing to that $\rho\geq 0$, a vector $|v\ra$ in
${\cal C}^2_{\sc A}\otimes{\cal C}^M_{\sc B}\otimes{\cal C}^N_{\sc C}$
satisfying $\la v|\rho|v\ra=0$ is in the kernel of $\rho$.
It is directly verified that the following vectors are in the kernel:
\be
\ba{ll}
|10\ra|f\ra-|13\ra
C|f\ra,&~~
|11\ra|g\ra-|13\ra B|g\ra,\\[3mm]
|12\ra|h\ra-|13\ra A|h\ra,&~~ |03\ra|k\ra-|13\ra D|k\ra,
\ea
\ee
for all $|f\ra,~|g\ra,~|h\ra,~|k\ra\in {\cal C}^N_{\sc C}$. This implies
\be\label{imp}
\ba{lll} E_{45}=D^{\dag}C,&~~E_{46}=D^{\dag}B,&~~
E_{47}=D^{\dag}A,\\[3mm]
E_{15}=E_{18}C,&~~
E_{25}=E_{28}C,&~~E_{35}=E_{38}C,\\[3mm]
E_{16}=E_{18}B,&~~ E_{26}=E_{28}B,&~~ E_{36}=E_{38}B,\\[3mm]
E_{17}=E_{18}A,&~~ E_{27}=E_{28}A,&~~ E_{37}=E_{38}A,\\[3mm]
E_{14}=E_{18}D,&~~ E_{24}=E_{28}D,&~~ E_{34}=E_{38}D.
\ea \ee

Substituting (\ref{imp}) into (\ref{rm}) and taking a partial
transposition of $\rho$ with respect to the sub-system ${\sc a}$,
we have \be \rho^{t_{\sc a}}=\left(
\begin{array}{cccccccc}
E_1&E_{12}&E_{13}&E_{18}D&C^{\dag}E_{18}^{\dag}
&C^{\dag}E_{28}^{\dag}&C^{\dag}E_{38}^{\dag}&C^{\dag}D\\
E_{12}^{\dag}&E_2&E_{23}&E_{28}D&
B^{\dag}E_{18}^{\dag}&B^{\dag}E_{28}^{\dag}&B^{\dag}E_{38}^{\dag}&B^{\dag}D\\
E_{13}^{\dag}&E_{23}^{\dag}&E_3&
E_{38}D&A^{\dag}E_{18}^{\dag}&A^{\dag}E_{28}^{\dag}&A^{\dag}E_{38}^{\dag}&A^{\dag}D\\
D^{\dag}E_{18}^{\dag}&D^{\dag}E_{28}^{\dag}&D^{\dag}E_{38}^{\dag}&D^{\dag}D&E_{18}^{\dag}
&E_{28}^{\dag}&E_{38}^{\dag}&D\\
E_{18}C&E_{18}B&E_{18}A&E_{18}&C^{\dag}C
&C^{\dag}B&C^{\dag}A&C^{\dag}\\
E_{28}C&E_{28}B&E_{28}A&E_{28}
&B^{\dag}C&B^{\dag}B&B^{\dag}A&B^{\dag}\\
E_{38}C&E_{38}B&E_{38}A&E_{38}
&A^{\dag}C&A^{\dag}B&A^{\dag}A&A^{\dag}\\
D^{\dag}C&D^{\dag}B&D^{\dag}A&D^{\dag}&C&B&A&I
\end{array}
\right).
\ee

Since the partial transposition with respect to the sub-system
${\sc a}$ is positive, $\rho^{t_{\sc a}}\ge 0$, and it does not
change $\la 1_{\sc a}|\rho|1_{\sc a}\ra$, we still have
$|10\ra|f\ra-|13\ra C|f\ra\in k(\rho^{t_{\sc a}})$. This gives
rise to the following equalities:
$E_{18}^{\dag}=DC,~E_{28}^{\dag}=DB,~E_{38}^{\dag}=DA$,
$E_{18}=C^{\dag}D^{\dag},~E_{28}=B^{\dag}D^{\dag},~E_{38}=A^{\dag}D^{\dag}$.
$\rho$ is then of the following form:
$$
\ba{l}
\left(
\begin{array}{cccccccc}
E_1\!&\!E_{12}\!&\!E_{13}\!&\!
C^{\dag}D^{\dag}D\!&\!C^{\dag}D^{\dag}C
\!&\!C^{\dag}D^{\dag}B\!&\!C^{\dag}D^{\dag}A
\!&\!C^{\dag}D^{\dag}\\
E_{12}^{\dag}&E_2&E_{23}&B^{\dag}D^{\dag}D\!&\!B^{\dag}D^{\dag}
C\!&\!B^{\dag}D^{\dag}B\!&\!B^{\dag}D^{\dag}A
\!&\!B^{\dag}D^{\dag}\\
E_{13}^{\dag}\!&\!E_{23}^{\dag}\!&\!E_3\!&\!
A^{\dag}D^{\dag}D\!&\!A^{\dag}D^{\dag}C\!&\!A^{\dag}D^{\dag}B\!&\!A^{\dag}D^{\dag}A
\!&\!A^{\dag}D^{\dag}\\
D^{\dag}DC\!&\!D^{\dag}DB\!&\!D^{\dag}DA\!&\!
D^{\dag}D\!&\!D^{\dag}C\!&\!D^{\dag}B\!&\!D^{\dag}A\!&\!D^{\dag}\\
C^{\dag}DC\!&\!C^{\dag}DB\!&\!C^{\dag}DA\!&\!
C^{\dag}D\!&\!C^{\dag}C\!&\!C^{\dag}B\!&\!C^{\dag}A
\!&\!C^{\dag}\\
B^{\dag}DC\!&\!B^{\dag}DB\!&\!B^{\dag}DA\!&\!
B^{\dag}D\!&\!B^{\dag}C\!&\!B^{\dag}B\!&\!B^{\dag}A\!&\!B^{\dag}\\
A^{\dag}DC\!&\!A^{\dag}DB\!&\!A^{\dag}DA\!&\!
A^{\dag}D\!&\!A^{\dag}C\!&\!A^{\dag}B\!&\!A^{\dag}A\!&\!A^{\dag}\\
DC\!&\!DB\!&\!DA\!&\!D\!&\!C\!&\!B\!&\!A\!&\!I
\end{array}
\right).\\[20mm]
\ea
$$
Set
$$
X=\left(
\begin{array}{cccccc}
E_{13}&C^{\dag}D^{\dag}D&C^{\dag}D^{\dag}C&C^{\dag}D^{\dag}B&C^{\dag}D^{\dag}A&C^{\dag}D^{\dag}\\
E_{23}&B^{\dag}D^{\dag}D&B^{\dag}D^{\dag}C&B^{\dag}D^{\dag}B&B^{\dag}D^{\dag}A&B^{\dag}D^{\dag}
\end{array}
\right),
$$
$$
Y=\left(
\begin{array}{cc}
E_1&E_{12}\\
E_{12}^{\dag}&E_2
\end{array}
\right),
$$
and
$$
\rho_6=\Sigma+{\rm diag}(\Delta,\ 0,\ 0,\ 0,\ 0,\ 0),
$$ \\
where $$ \Sigma=\left(
\begin{array}{cccccc}
A^{\dag}D^{\dag}DA&A^{\dag}D^{\dag}D&A^{\dag}
D^{\dag}C&A^{\dag}D^{\dag}B&A^{\dag}D^{\dag}A&A^{\dag}D^{\dag}\\
D^{\dag}DA&D^{\dag}D&D^{\dag}C&D^{\dag}B&D^{\dag}A&D^{\dag}\\
C^{\dag}DA&C^{\dag}D&C^{\dag}C&C^{\dag}B&C^{\dag}A&C^{\dag}\\
B^{\dag}DA&B^{\dag}D&B^{\dag}C&B^{\dag}B&B^{\dag}A&B^{\dag}\\
A^{\dag}DA&A^{\dag}D&A^{\dag}C&A^{\dag}B&A^{\dag}A&A^{\dag}\\
DA&D&C&B&A&I
\end{array}
\right),
$$
\be\label{delta} \Delta=E_3-A^{\dag}D^{\dag}DA, \ee
diag$(A_1,A_2,...,A_m)$ denotes a diagonal block matrix with
blocks $A_1,A_2,...,A_m$. $\rho$ can then be written in the
following partitioned matrix form:
\[
\rho=\left(
\begin{array}{cc}
Y&X\\
X^{\dag}&\rho_6
\end{array}
\right).
\]
As $\Sigma$ possesses the following  5N kernel vectors: \be
\ba{ll} (\la f|, \ 0,\ 0,\ 0,\ 0, -\la f|A^\dag D^\dag  )^T,\ \
&~~~
(0, \la g|,0,\ 0,\ 0, -\la g|D^\dag )^T,\\[3mm]
(0,\ 0,\la h|, \ 0,\ 0,\ -\la h|C^\dag )^T,\ \ &~~~
(0,\ 0,\ 0,\la i|,\ 0,\ -\la i|B^\dag )^T,\\[3mm]
\ \ (0,\ 0,\ 0,\ 0,\la j|,\ -\la j|A^\dag )^T, \ea \ee for
arbitrary $|f\ra,\ |g\ra,\ |h\ra,\ |i\ra,\ |j\ra \in{\cal C}^N_{\sc C}$,
so the kernel $K(\Sigma)$ has at least dimension $5N$. On
the other hand $r(\Sigma)+k(\Sigma)=6N$, therefore $r(\Sigma)\leq
N$. While the range of $\Sigma$ has at least dimension $N$ due to
the identity entry on the diagonal. Hence we have $r(\Sigma)=N$.

Taking into account that $r(\rho_6)\leq r(\rho)=N$, it is easy to see that
$r(\rho_6)=N$. As the rank of $\rho_6$ is $N$, by making the following
elementary row transformations on the matrix $\rho_6$,
\be\label{add} \left(
\begin{array}{cccccc}
I&0&0&0&0&-A^{\dag}D^{\dag}\\
0&I&0&0&0&-D^{\dag}\\
0&0&I&0&0&-C^{\dag}\\
0&0&0&I&0&-B^{\dag}\\
0&0&0&0&I&-A^{\dag}\\
0&0&0&0&0&I
\end{array}
\right)\rho_6= \left(
\begin{array}{cccccc}
\Delta&0&0&0&0&0\\
0&0&0&0&0&0\\
0&0&0&0&0&0\\
0&0&0&0&0&0\\
0&0&0&0&0&0\\
DA&D&C&B&A&I
\end{array}
\right). \ee we deduce that $\Delta=0$, i.e.,
$E_3=A^{\dag}D^{\dag}DA$.

Similarly, $|02\ra |f\ra -|13\ra DA|f\ra$
is also in the kernel for all $|f\ra\in{\cal C}^N_{\sc C}$,
from which we have $E_{13}=C^{\dag}D^{\dag}DA$,
$E_{23}=B^{\dag}D^{\dag}DA$ and the matrix $\rho$ becomes
$$
\left(
\begin{array}{cccccccc}
E_1\!&\!E_{12}\!&\! C^{\dag}D^{\dag}DA\!&\!
C^{\dag}D^{\dag}D\!&\!C^{\dag}D^{\dag}C
\!&\!C^{\dag}D^{\dag}B\!&\!C^{\dag}D^{\dag}A
\!&\!C^{\dag}D^{\dag}\\
E_{12}^{\dag}&E_2&B^{\dag}D^{\dag}DA&B^{\dag}D^{\dag}D\!&\!
B^{\dag}D^{\dag}C\!&\!B^{\dag}D^{\dag}B\!&\!B^{\dag}D^{\dag}A
\!&\!B^{\dag}D^{\dag}\\
A^{\dag}D^{\dag}DC\!&\!A^{\dag}D^{\dag}DB\!&\!A^{\dag}D^{\dag}DA\!&\!
A^{\dag}D^{\dag}D\!&\!A^{\dag}D^{\dag}C\!&\!A^{\dag}D^{\dag}B\!&\!A^{\dag}D^{\dag}A
\!&\!A^{\dag}D^{\dag}\\
D^{\dag}DC\!&\!D^{\dag}DB\!&\!D^{\dag}DA\!&\!
D^{\dag}D\!&\!D^{\dag}C\!&\!D^{\dag}B\!&\!D^{\dag}A\!&\!D^{\dag}\\
C^{\dag}DC\!&\!C^{\dag}DB\!&\!C^{\dag}DA\!&\!
C^{\dag}D\!&\!C^{\dag}C\!&\!C^{\dag}B\!&\!C^{\dag}A
\!&\!C^{\dag}\\
B^{\dag}DC\!&\!B^{\dag}DB\!&\!B^{\dag}DA\!&\!
B^{\dag}D\!&\!B^{\dag}C\!&\!B^{\dag}B\!&\!B^{\dag}A\!&\!B^{\dag}\\
A^{\dag}DC\!&\!A^{\dag}DB\!&\!A^{\dag}DA\!&\!
A^{\dag}D\!&\!A^{\dag}C\!&\!A^{\dag}B\!&\!A^{\dag}A\!&\!A^{\dag}\\
DC\!&\!DB\!&\!DA\!&\!D\!&\!C\!&\!B\!&\!A\!&\!I
\end{array}
\right).\\
$$
Similarly, using the same method  as for $E_3$ and $E_{23}$
respectively, we can derive $E_2=B^{\dag}D^{\dag}DB$,
$E_{12}=C^{\dag}D^{\dag}DB$, $E_1=C^{\dag}D^{\dag}DC$. $\rho$ then
is the following form:
$$
\ba {l} \left(
\begin{array}{cccccccc}
C^{\dag}D^{\dag}DC\!&\!C^{\dag}D^{\dag}DB\!&\!C^{\dag}D^{\dag}DA\!&\!
C^{\dag}D^{\dag}D\!&\!C^{\dag}D^{\dag}C
\!&\!C^{\dag}D^{\dag}B\!&\!C^{\dag}D^{\dag}A
\!&\!C^{\dag}D^{\dag}\\
B^{\dag}D^{\dag}DC&B^{\dag}D^{\dag}DB&B^{\dag}D^{\dag}DA&B^{\dag}D^{\dag}
D\!&\!B^{\dag}D^{\dag}C\!&\!B^{\dag}D^{\dag}B\!&\!B^{\dag}D^{\dag}A
\!&\!B^{\dag}D^{\dag}\\
A^{\dag}D^{\dag}DC\!&\!A^{\dag}D^{\dag}DB\!&\!A^{\dag}D^{\dag}DA\!&\!
A^{\dag}D^{\dag}D\!&\!A^{\dag}D^{\dag}C\!&\!A^{\dag}D^{\dag}B\!&\!A^{\dag}D^{\dag}A
\!&\!A^{\dag}D^{\dag}\\
D^{\dag}DC\!&\!D^{\dag}DB\!&\!D^{\dag}DA\!&\!
D^{\dag}D\!&\!D^{\dag}C\!&\!D^{\dag}B\!&\!D^{\dag}A\!&\!D^{\dag}\\
C^{\dag}DC\!&\!C^{\dag}DB\!&\!C^{\dag}DA\!&\!
C^{\dag}D\!&\!C^{\dag}C\!&\!C^{\dag}B\!&\!C^{\dag}A
\!&\!C^{\dag}\\
B^{\dag}DC\!&\!B^{\dag}DB\!&\!B^{\dag}DA\!&\!
B^{\dag}D\!&\!B^{\dag}C\!&\!B^{\dag}B\!&\!B^{\dag}A\!&\!B^{\dag}\\
A^{\dag}DC\!&\!A^{\dag}DB\!&\!A^{\dag}DA\!&\!
A^{\dag}D\!&\!A^{\dag}C\!&\!A^{\dag}B\!&\!A^{\dag}A\!&\!A^{\dag}\\
DC\!&\!DB\!&\!DA\!&\!D\!&\!C\!&\!B\!&\!A\!&\!I
\end{array}
\right)\\[20mm]
=[\begin{array}{cccccccc} DC& DB&DA&D&C&B&A&I\end{array}]^{\dag}
[DC\,\,\, DB\,\,\, DA\,\,\, D\,\,\, C\,\,\, B\,\,\, A\,\,\, I].\\[3mm]
\ea
$$

The commutative relations $[A,\ D]=[B,\ D]=[C,\ D]=[A,\
D^{\dag}]=[B,\ D^{\dag}]=[C,\ D^{\dag}]=0$ follow from the
positivity of all partial transpositions of $\rho$. We first
consider: \be \rho^{t_{\sc b}}=\left(
\begin{array}{cccccccc}
C^{\dag}D^{\dag}DC\!&\!B^{\dag}D^{\dag}DC\!&\!A^{\dag}D^{\dag}DC\!&\!
D^{\dag}DC\!&\!C^{\dag}D^{\dag}C
\!&\!B^{\dag}D^{\dag}C\!&\!A^{\dag}D^{\dag}C
\!&\!D^{\dag}C\\
C^{\dag}D^{\dag}DB\!&\!B^{\dag}D^{\dag}DB\!&\!A^{\dag}D^{\dag}DB\!&\!
D^{\dag}DB\!&\!C^{\dag}D^{\dag}B\!&\!B^{\dag}D^{\dag}B\!&\!A^{\dag}D^{\dag}B
\!&\!D^{\dag}B\\
C^{\dag}D^{\dag}DA\!&\!B^{\dag}D^{\dag}DA\!&\!A^{\dag}D^{\dag}DA\!&\!
D^{\dag}DA\!&\!C^{\dag}D^{\dag}A\!&\!B^{\dag}D^{\dag}A\!&\!A^{\dag}D^{\dag}A
\!&\!D^{\dag}A\\
C^{\dag}D^{\dag}D\!&\!B^{\dag}D^{\dag}D\!&\!A^{\dag}D^{\dag}D\!&\!
D^{\dag}D\!&\!C^{\dag}D^{\dag}\!&\!B^{\dag}D^{\dag}\!&\!A^{\dag}D^{\dag}\!&\!D^{\dag}\\
C^{\dag}DC\!&\!B^{\dag}DC\!&\!A^{\dag}DC\!&\!
DC\!&\!C^{\dag}C\!&\!B^{\dag}C\!&\!A^{\dag}C
\!&\!C\\
C^{\dag}DB\!&\!B^{\dag}DB\!&\!A^{\dag}DB\!&\!
DB\!&\!C^{\dag}B\!&\!B^{\dag}B\!&\!A^{\dag}B\!&\!B\\
C^{\dag}DA\!&\!B^{\dag}DA\!&\!A^{\dag}DA\!&\!
DA\!&\!C^{\dag}A\!&\!B^{\dag}A\!&\!A^{\dag}A\!&\!A\\
C^{\dag}D\!&\!B^{\dag}D\!&\!A^{\dag}D\!&\!D\!&\!C^{\dag}\!&\!B^{\dag}\!&\!A^{\dag}\!&\!I
\end{array}
\right).
\ee
Due to the positivity, the matrix $\rho^{t_B}$ must
possess the kernel vector $|03\ra |h\ra -|13\ra D |h\ra$, which
implies that $[A,\ D]=[B,\ D]=[C,\ D]=0$.The matrix $\rho^{t_B}$
can be then written as:
\[
\rho^{t_B}=\left(
\begin{array}{c}
D^{\dag}C\\
D^{\dag}B\\
D^{\dag}A\\
D^{\dag}\\
C\\
B\\
A\\
I
\end{array}
\right) \left(
\begin{array}{cccccccc}
C^{\dag}D&B^{\dag}D&A^{\dag}D&D&C^{\dag}&B^{\dag}&A^{\dag}&I
\end{array}
\right),
\]
which implies automatically the positivity.

From the positivity of $\rho^{t_{AB}}$,
 \[
\rho^{t_{AB}}=\left(
\begin{array}{cccccccc}
C^{\dag}D^{\dag}DC&B^{\dag}D^{\dag}DC&A^{\dag}D^{\dag}DC&D^{\dag}DC&C^{\dag}DC&B^{\dag}DC&A^{\dag}DC&DC\\
C^{\dag}D^{\dag}DB&B^{\dag}D^{\dag}DB&A^{\dag}D^{\dag}DB&D^{\dag}DB&C^{\dag}DB&B^{\dag}DB&A^{\dag}DB&DB\\
C^{\dag}D^{\dag}DA&B^{\dag}D^{\dag}DA&A^{\dag}D^{\dag}DA&D^{\dag}DA&C^{\dag}DA&B^{\dag}DA&A^{\dag}DA&DA\\
C^{\dag}D^{\dag}D&B^{\dag}D^{\dag}D&A^{\dag}D^{\dag}D&D^{\dag}D&C^{\dag}D&B^{\dag}D&A^{\dag}D&D\\
C^{\dag}D^{\dag}C&B^{\dag}D^{\dag}C&A^{\dag}D^{\dag}C&D^{\dag}C&C^{\dag}C&B^{\dag}C&A^{\dag}C&C\\
C^{\dag}D^{\dag}B&B^{\dag}D^{\dag}B&A^{\dag}D^{\dag}B&D^{\dag}B&C^{\dag}B&B^{\dag}B&A^{\dag}B&B\\
C^{\dag}D^{\dag}A&B^{\dag}D^{\dag}A&A^{\dag}D^{\dag}A&D^{\dag}A&C^{\dag}A&B^{\dag}A&A^{\dag}A&A\\
 C^{\dag}D^{\dag}&B^{\dag}D^{\dag}&A^{\dag}D^{\dag}&D^{\dag}&C^{\dag}&B^{\dag}&A^{\dag}&I
\end{array}
\right),
\]
we have that $|03\ra |h\ra -|13\ra D^{\dag} |h\ra$ is a kernel
vector, which results in $[A,\ D^{\dag}]=[B,\ D^{\dag}]=[C,\
D^{\dag}]=0$. $\rho^{t_{AB}}$ is then of the form:
\[
\rho^{t_{AB}}=\left(
\begin{array}{c}
DC\\
DB\\
DA\\
D\\
C\\
B\\
A\\
I
\end{array}
\right) \left(
\begin{array}{cccccccc}
C^{\dag}D^{\dag}&B^{\dag}D^{\dag}&A^{\dag}D^{\dag}&D^{\dag}&C^{\dag}&B^{\dag}&A^{\dag}&I
\end{array}
\right).
\]
This form assures positive definiteness, and concludes the proof
of the Lemma. $\Box$

Using Lemma 1 we can prove the following Theorem:

{\bf Theorem 1.}\ \ A PPT-state $\rho$ in ${\cal C}^2_{\sc A}
\otimes {\cal C}^4_{\sc B} \otimes {\cal C}^N_{\sc C}$ with
$r(\rho)=N$ is separable if there exists a product vector $|e_A,\
f_B\ra$ of ${\cal C}^2_{\sc A} \otimes {\cal C}^4_{\sc B}$ such
that $r(\la e_A,\ f_B|\rho |e_A,\ f_B\ra)=N$.

{\bf Proof.}\ \ According to the Lemma 1 the PPT state $\rho$ can
be written as
\[
\rho=
\left(
\begin{array}{c}
C^{\dag}D^{\dag}\\B^{\dag}D^{\dag}\\A^{\dag}D^{\dag}\\
D^{\dag}\\C^{\dag}\\B^{\dag}\\A^{\dag}\\I
\end{array}
\right)
\left(
\begin{array}{cccccccc}
DC&DB&DA&D&C&B&A&I
\end{array}
\right).
\]
Since all $A$, $A^\dag$, $B$, $B^\dag$, $C$, $C^\dag$, $D$ and
$D^\dag$ commute, they have common eigenvectors $|f_n\ra$. Let
$a_n$, $b_n$, $c_n$ and $d_n$ be the corresponding eigenvalues of
$A$, $B$, $C$ and $D$ respectively. We have
\[ \la f_n|\rho
|f_n\ra= \left(
\begin{array}{c}
c_n^*d_n^*\\[2mm]b_n^*d_n^*\\[2mm]a_n^*d_n^*\\[2mm]d_n^*\\[2mm]c_n^*\\[2mm]
b_n^*\\[2mm]a_n^*\\[2mm]1
\end{array}
\right)
\left(
\begin{array}{cccccccc}
d_nc_n&d_nb_n&d_na_n&d_n&c_n&b_n&a_n&1
\end{array}
\right)\hskip5cm
\]
\[
=\left[\left(\begin{array}{c}d_n^*\\1\end{array}\right)\otimes
\left(\begin{array}{c}c_n^*\\b_n^*\\a_n^*\\1\end{array}\right)\right]
(d_n\,\,\, 1)\otimes(c_n,b_n,a_n,\,1)
=|e_{\sc a},f_{\sc b}\ra\la
e_{\sc a},f_{\sc b}|.
\]
We can thus write $\rho$ as
$$\rho=\sum_{n=1}^N|\psi_n\ra\la \psi_n|\otimes |\phi_n\ra\la \phi_n|
\otimes |f_n\ra\la f_n|,
$$
where
$$
|\psi_n\ra=\left(\begin{array}{c}d_n^*\\ 1\end{array}\right),~~~
|\phi_n\ra=\left(\begin{array}{c}c_n^*\\b_n^*\\a_n^*\\
 1\end{array}\right).
$$
Since the local transformations are reversible, we can apply the
inverse transformations and obtain a decomposition of the initial
state $\rho$ in a sum of projectors onto the product vectors.
Therefore $\rho$ is separable. \hfill $\Box$

The above approach can be extended to the higher dimensional case
${\cal C}^2_{\sc a} \otimes {\cal C}^M_{\sc b}\otimes {\cal
C}^N_{\sc c}$. Similar to Lemma 1, it is straightforward to prove
the following conclusion:

 {\bf Lemma 2. }\ \  Every PPT state $\rho$ in
${\cal C}^2_{\sc a} \otimes {\cal C}^M_{\sc b}\otimes {\cal
C}^N_{\sc c}$ such that $r(\la 1_A, M-1_B|\rho |1_A,
M-1_B\ra)=r(\rho)=N$, can be transformed into the following
canonical form by using a reversible local operation:
$$
\rho=\sqrt{F}[DA_{M-1}\ \cdots\ \ DA_1\ \ D \ \ A_{M-1}\ \cdots\ \
A_1\ \ I]^{\dag} [DA_{M-1}\ \cdots\ \ DA_1\ \ D \ \ A_{M-1}\
\cdots\ \ A_1\ \ I]\sqrt{F},
$$
where $A_1$, $\cdots$, $A_{M-1}$, $D$, $F$ and the identity $I$
are $N\times N$ matrices acting on ${\cal C}_{\sc c}^N$ and
satisfy the following relations: $[A_i,\ {A_j}^{\dag}]=[A_i,\
A_j]=[D,\ D^{\dag}]=[D,\ A_i]=[D,\ {A_i}^{\dag}]=0$,
$i,j=1,2,\cdots,M-1$ and $F=F^{\dag}$ . Set $T=(D\ \
I)\otimes(A_{M-1}\ \ A_{M-2}\ \cdots\ \ A_1\ \ I)$, we have \be
\rho=\sqrt{F} T^{\dag}T \sqrt{F}\ee

Extending Theorem 1 to higher dimensional cases
${\cal C}^2_{\sc a} \otimes {\cal C}^M_{\sc
b}\otimes {\cal C}^N_{\sc c}$, we have:

{\bf Theorem 2.}\ \ A PPT-state $\rho$ in ${\cal C}^2 \otimes
{\cal C}^M \otimes {\cal C}^N$ with $r(\rho)=N$ is separable if
there exists a product vector $|e_A,\ f_B\ra$ such that $r(\la
e_A,\ f_B|\rho |e_A,\ f_B\ra)=N$.

A separable state is always PPT. But a PPT state in ${\cal C}^2
\otimes {\cal C}^M \otimes {\cal C}^N$ is not generally separable.
Our canonical form of PPT states in ${\cal C}^2 \otimes {\cal C}^M
\otimes {\cal C}^N$ composite quantum systems provides a way to
investigate the separability of such states and the structures of
bound entangled states, as all entangled PPT states are bound
entangled states.

\end{document}